\documentclass[aps, prl, reprint, twocolumn, longbibliography]{revtex4-2}
\usepackage[T1]{fontenc}           
\usepackage[british]{babel} 
\usepackage{amsmath}
\usepackage{amsfonts}
\usepackage{mathtools}
\usepackage{color}
\usepackage{graphicx} % Required for inserting images
\usepackage{hyperref}
\usepackage{comment}
\usepackage{braket}
\usepackage{fullpage}
\DeclareMathOperator{\Tr}{Tr}
\renewcommand{\Re}{\mathop{\mathrm{Re}}\nolimits}
\renewcommand{\Im}{\mathop{\mathrm{Im}}\nolimits}

\begin{document}

\title{Nonclassical Photon-Assisted Transport in Superconducting Tunnel Junctions}
\author{Matthias H\"ubler}
\affiliation{Fachbereich Physik, Universit\"at Konstanz, 78457 Konstanz, Germany}

\author{Juan Carlos Cuevas}
\affiliation{Departamento de Física Teórica de la Materia Condensada, Universidad Autónoma de Madrid, E-28049 Madrid, Spain}
\affiliation{Condensed Matter Physics Center (IFIMAC), Universidad Aut\'{o}noma de Madrid, E-28049 Madrid, Spain}

\author{Wolfgang Belzig}
\affiliation{Fachbereich Physik, Universit\"at Konstanz, 78457 Konstanz, Germany}

\date{\today}

\begin{abstract}
Advances in circuit quantum electrodynamics have enabled the generation of arbitrary nonclassical microwave states and
paved the way for addressing novel physics questions. Here, we present a theoretical study of the electrical current
in a Josephson tunnel junction interacting with a nonclassical electromagnetic environment. This allows us to 
generalize classical transport phenomena like photon-assisted tunneling and Shapiro steps to the quantum regime. We
predict that the analysis of the supercurrent in such a setup enables the complete reconstruction of quantum states 
of the electromagnetic environment, something that is not possible with normal tunnel junctions.
\end{abstract}

\maketitle

\textit{Introduction.---} The first investigations of microwave-irradiated Josephson tunnel junctions date back to the early 
1960’s \cite{Barone1982},  which in particular revealed two textbook phenomena. First, the appearance of microwave-induced
steps in the current-voltage characteristics \cite{Dayem1962}, which were interpreted as \emph{photon-assisted tunneling} 
of single quasiparticles \cite{Tien1963}. Second, the occurrence of \emph{Shapiro steps}, which consists of
additional contributions to the dc current at very specific bias voltages that stem from the phase locking between the 
ac Josephson current components and the microwave field \cite{Shapiro1963}. These two phenomena illustrated the duality that Josephson junctions can be used as microwave detectors and microwaves can be used to learn about the Josephson effect. With the advent of circuit quantum electrodynamics (circuit QED) in recent years \cite{Blais2021,Clerk2021}, 
and the ability to create nonclassical microwave states almost at will \cite{Hofheinz2009}, a natural question arises: 
what novel physical phenomena may result when Josephson tunnel junctions are subjected to nonclassical microwaves?

In fact, the nonclassical microwaves generated by superconducting resonators have already provided novel insight 
into the physics of Josephson junctions via microwave spectroscopy of Andreev bound states, see e.g.
\cite{Bretheau2013,Bretheau2014}. It has also been shown that the coupling of a high impedance microwave resonator to a 
Josephson tunnel junction can be used for quantum bath engineering \cite{Aiello2022}. 

In this Letter, we address the question above and present a theoretical study of the current-voltage characteristics of 
Josephson tunnel junctions interacting with nonclassical microwave states. Our main goal is to show that superconducting 
tunnel junctions can be used as detectors to reveal the nature of nonclassical microwave states beyond the capabilities 
of normal (nonsuperconducting) tunnel junctions \cite{Souquet2014}. To be precise, we show how the analysis of the 
supercurrent and the quasiparticle current flowing through the Josephson junction enables the complete reconstruction of 
the density matrix of quantum states fabricated with circuit QED setups. 

\begin{figure}
    \centering
    \includegraphics[width=1\linewidth]{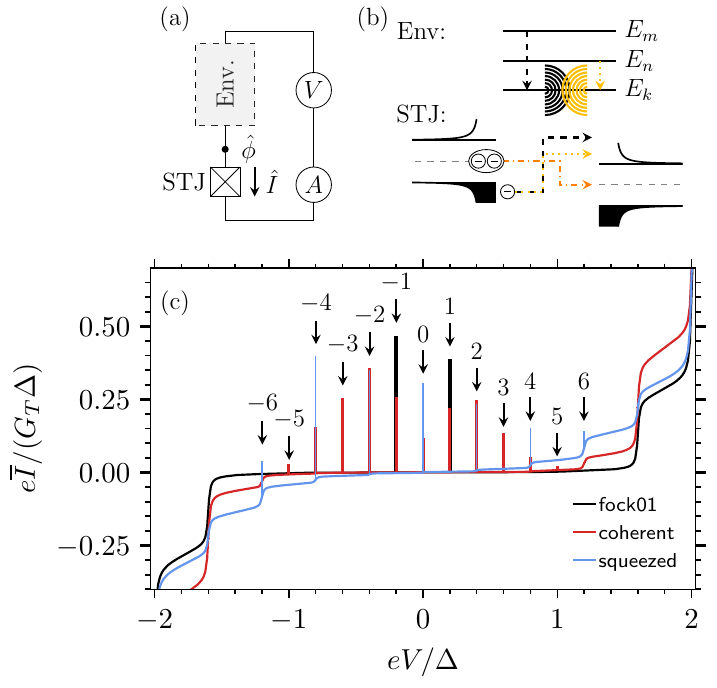}
    \caption{(a) Superconducting tunnel junction (STJ) interacting with an electromagnetic environment (Env.)\
    that imposes a quantum phase $\hat{\phi}$ on the junction. A DC voltage $V$ is applied, and the current-voltage
    characteristic is investigated. (b) Transitions $E_m \rightarrow E_k$ or $E_n \rightarrow E_k,\ldots$ between the 
    discrete energy levels of the electromagnetic environment induce quasiparticle transport in the STJ. Interference
    between a pair of transitions leads to inelastic Cooper pair transport, which manifests as current spikes at 
    $eV=(E_m-E_n)/2$ in the DC current. (c) DC current as a function of the voltage for an STJ at 
    zero temperature, coupled to an $LC$-resonator with frequency $\hbar \omega_0=0.4 \Delta$ and coupling strength 
    $g=1/2$. The oscillator is in the states defined in Eqs.~\eqref{eq:states}, where $|\xi|=|\alpha|=1$, 
    and the phases are arbitrary. The spikes represent the amplitude of the Cooper pair current-phase relation
    \eqref{eq:CPosci}, while the background corresponds to the quasiparticle current \eqref{eq:QPOsci}. 
    Spikes appear at $eV=j \hbar \omega_0/2$ (indicated by the arrows), whereas steps occur at $eV=2\Delta+l\hbar 
    \omega_0$, with $j,l\in \mathbb{Z}$.}
    \label{fig:Opener}
\end{figure}

\textit{General environment.---} We investigate how a general electromagnetic environment modifies quantum transport in a superconducting tunnel junction (see Fig.~\ref{fig:Opener}(a) for the setup). 
The environment imposes a voltage, and thereby a dynamical phase 
$\hat{\phi}(t)$, across the junction. This phase couples to the tunneling current
$\hat{I}$ via the interaction term $\hat{H}_{\text{int}}(t)=(\hbar/e) \,\hat{\phi}(t) \hat{I}$, where $\hbar$ is the reduced Planck constant, and $e$ is the elementary
charge. For example, a constant voltage bias $V$ induces the phase $\hat{\phi}
(t)=\varphi_{\text{dc}}(t)=eV (t-t_s)/\hbar+\phi_0/2$, with $\phi_0/2$ the phase at the initial time $t_s$. Photon-assisted tunneling is described by an 
additional phase $\varphi(t)=\alpha \sin(\omega_0 (t-t_s) )$, with $\alpha$ the strength of the classical drive, and
$\omega_0$ the driving frequency~\cite{Tien1963,Barone1982}. A quantum environment introduces a phase
operator $\hat{\varphi}$, whose dynamics is governed by an environmental
Hamiltonian $\hat{H}_\text{env}$. In the interaction picture, this
phase evolves as $\hat{\varphi}(t)=\exp(i \hat{H}_\text{env}(t-t_s))\hat{\varphi}\exp(-i \hat{H}_\text{env}(t-t_s))$ 
leading to a total phase across the junction given by $\hat{\phi}(t)=\hat{\varphi}(t)+\varphi_{\text{dc}}(t)$~\cite{SupMat}. 

Our goal is the analysis of the tunnel current in a superconducting junction
arising from both a voltage bias and an arbitrary quantum electromagnetic
environment (see Fig.~\ref{fig:Opener}(a)). We employ the Keldysh
formalism~\cite{Nazarov2012} to compute the cumulant-generating functional of the hybrid system to 
the lowest order in the tunnel conductance $G_\text{T}$~\cite{SupMat}. 
%This formalism has been used to treat mesoscopic conductors under arbitrary classical drives~\cite{Vanevic2007,Vanevic2008}, 
%as well as electromagnetic environments in equilibrium~\cite{Kindermann2003,Kindermann2004,Tobiska2005}.
In principle, the current, noise, and higher-order cumulants can be obtained from the cumulant-generating function. 
We focus here on the tunnel current, as it already captures a broad range of physical effects.
The tunnel current comprises quasiparticle (qp) and Cooper pair (cp) transport, and takes the form:
\begin{subequations}
\begin{align}
I(t)=& I_{\text{qp}}(t)+I_{\text{cp}}(t) \nonumber\\
=&\sum_{\nu=\pm } \nu \int \limits_{-\infty}^\infty dE \, P^\nu(t,E) \Gamma_{\text{qp}}(E) \label{eq:QPCurr}\\
+&\sum_{\nu=\pm } \nu \int \limits_{-\infty}^\infty dE \, C^\nu(t,E) \Gamma_{\text{cp}}(E). \label{eq:CPCurr}
\end{align}
\end{subequations}
Both contributions depend on the tunneling rate of the isolated junction, which are given by
\begin{subequations}
\begin{align}
    \Gamma_{\text{qp}}(E)=&\frac{G_T}{e} \hspace{-1pt}\int \limits_{-\infty}^\infty d \epsilon  \varrho_1(\epsilon-E) 
    \varrho_2(\epsilon) [1-f(\epsilon-E)] f(\epsilon) \label{eq:QPTunnelrate}\\
    \Gamma_{\text{cp}}(E)=&\frac{G_T}{e} \hspace{-1pt} \int \limits_{-\infty}^\infty  d \epsilon  \varsigma_1(\epsilon-E) 
    \varsigma_2(\epsilon) [1-f(\epsilon-E)] f(\epsilon), \label{eq:CPTunnelrate}
\end{align}
\end{subequations}
with the normalized quasiparticle density of states $\varrho_p(\epsilon) = \Theta(|\epsilon|-|\Delta_p|) 
|\epsilon|/\sqrt{\epsilon^2-\Delta_p^2}$, the normalized pair density $\varsigma_p(\epsilon) = 
\Theta(|\epsilon|-|\Delta_p|)|\Delta_p|/\sqrt{\epsilon^2-\Delta_p^2}$, the superconducting gap $\Delta_p\equiv\Delta$ of terminal
$p=1,2$, and the Fermi-distribution $f(\epsilon)=[1+\exp(\epsilon/(k_\text{B}T))]^{-1}$ at thermal energy 
$k_\text{B}T$. 
Here, the terminals are assumed to be s-wave superconductors.
The $\nu=+$ terms in Eqs.~\eqref{eq:QPCurr} and \eqref{eq:CPCurr} describe transport from terminal 
$2$ to terminal $1$, while the $\nu=-$ terms describe transport in the opposite direction. The environment 
influences the transport via the generalized $P$-functions~\cite{Souquet2014} 
\begin{subequations}\begin{align}
P^\nu(t,E)=& \int \limits_{-\infty}^{\infty} d \tau \Im \bigg ( e^{i E \tau/\hbar} G^{\nu+}_{\text{env}}(t,\tau) 
\bigg ) \label{eq:Pfunc}
\end{align}
and $C$-functions
\begin{align}
C^\nu(t,E)=& \int \limits_{-\infty}^{\infty} d \tau \Im \bigg (e^{iE \tau/\hbar}  G^{\nu-}_{\text{env}}(t,\tau) 
\bigg ), 
\end{align}
which depend on the environmental correlator
\begin{align}
    G^{\nu\mu}_{\text{env}}(t,\tau)=&\frac{i}{\pi \hbar}\Theta(\tau) \bigg \langle e^{i\mu \nu \hat{\phi}(t)}
    e^{-i\nu \hat{\phi}(t-\tau)}\bigg \rangle. 
\end{align}\label{eq:Envcorr}\end{subequations}

The structure of the quasiparticle current $I_{\text{qp}}(t)$ is 
identical to that found in a normal-metal tunnel junction coupled to an arbitrary electromagnetic 
environment~\cite{Souquet2014}. The only difference is that the quasiparticle rates involve the superconducting rather 
than the normal-metal density of states. The $P$-functions in Eq.~\eqref{eq:Pfunc} coincide with that of the normal-metal
case~\cite{Souquet2014} and reduce to the known result from dynamical Coulomb blockade theory for an environment in 
equilibrium~\cite{Ingold1992,Joyez2013,Grabert2015}. The key difference between a quantum and a classical environment is 
that phase operators at different times do not commute, leading to distinct $P$-functions ($\nu=\pm$) for charge
transport from terminal $1$ to $2$ versus from $2$ to $1$. If the phase operator commutes at all times, the symmetry 
$P^-(t,E)=P^+(t,-E)$ is recovered. This symmetry ensures that the current depends directly on the bare current,
$I_{0,\text{qp}}(E)=\Gamma_{\text{qp}}(-E)-\Gamma_{\text{qp}}(E)$, as is the case in photon-assisted 
tunneling~\cite{Tien1963,Shapiro1963,Barone1982}. In a normal-metal junction, this non-commutativity leads to a 
nonlinear $I$-$V$ characteristic, as opposed to the linear behavior predicted by photon-assisted tunneling, which 
is attributed to the zero-point fluctuations of the field~\cite{Souquet2014}.

The Cooper pair current $I_{\text{cp}}(t)$ has a similar structure, with the density of states 
replaced by the pair density, and with the exponents in the environmental correlator carrying the same sign.
This sign difference profoundly affects the system's behavior and is responsible for both the DC and 
AC Josephson effects. The standard DC Josephson effect is recovered when the phase is constant, resulting in 
a sinusoidal current-phase relation. A DC bias induces a linear time-dependence of the phase,
leading to an oscillating Cooper pair current at the Josephson frequency, as can be derived from 
Eqs.~\eqref{eq:Envcorr}~\cite{SupMat}.

\textit{Discrete spectrum.---} We focus now on a broad class of environments composed of discrete modes. This 
class is encountered in circuit QED systems~\cite{Blais2021}, where the environment may
consist of an $LC$-resonator, an artificial atom, a superconducting qubit, etc. The 
environmental Hamiltonian $\hat{H}_{\text{env}}$ has a discrete spectrum, with eigenenergies $E_n$ 
and eigenstates $\ket{E_n}$. The phase $\hat{\varphi}(t)$ evolves according to the unitary operator $\hat{U}(t,t_s)$,
which takes the spectral form $\hat{U}(t,t_s)=\sum_n \exp(-i E_n(t-t_s)) \ket{E_n}\bra{E_n}$, where the sum runs 
over all eigenstates. We consider the electromagnetic environment to be initialized in a general quantum state,
described by the density matrix $\hat{\rho}$. The environmental correlator reduces to
\begin{align}
\langle e^{-i \nu \hat{\varphi}(t)} e^{-i \nu' \hat{\varphi}(t-\tau)} \rangle=&\sum_{n,m,k} \langle \hat{L}_{mk} 
\hat{L}_{kn} \rangle A^\nu_{mk} A^{\nu'}_{kn} \nonumber \\
\times&   e^{-i(E_k-E_n)\tau/\hbar-i (E_n-E_m)t/\hbar }, 
\end{align}
with the ladder operator $\hat{L}_{kn}=\ket{E_k}\bra{E_n}$ corresponding to the transition $E_n \rightarrow E_k$, 
and the transition amplitude $A^\nu_{kn}=\bra{E_k}\exp(-i\nu \hat{\varphi})\ket{E_n}$. The operator 
$\exp(-i \nu \hat{\varphi})$ acts as a translation operator for the charge operator $\hat{Q}$, which is 
canonically conjugate to the phase $\hat{\varphi}$, shifting the charge in the environment by one elementary 
unit~\cite{Ingold1992}.

The current $I(t)$ oscillates in time, and we focus on its DC component, given by the long-time average 
$\overline{I}=\lim_{T\rightarrow \infty}\int_{-T/2}^{T/2} I(t) dt$. The quasiparticle current is a superposition
of oscillations at transition frequencies between the states, and the time-averaged current is given by
\begin{subequations}\begin{align}
\overline{I}_{\text{qp}}=&  \sum_{\nu=\pm} \nu \sum_{n,k} \gamma^\nu[k,n]  \Gamma_{\text{qp}}(E_k-E_n-\nu eV),
\end{align}
where 
\begin{align}
    \gamma^\nu[k,n]=& |A^\nu_{kn}|^2 \Tr \{ \hat{L}_{kn} \hat{\rho} \hat{L}^\dagger_{kn} \}=|A^\nu_{kn}|^2 
    \rho_{nn} .
\end{align}\end{subequations}
Here, $\rho_{nn}$ are the diagonal entries of the density matrix in the energy basis and $\Tr \{ \cdot \}$ represents 
the trace. The transition rate $\gamma^\nu[k,n]$ from state $E_n$ to state $E_k$ involves the transition probability
$|A^\nu_{kn}|^2$ induced by the charge transport and the occupation probability of the initial state $\rho_{nn}$. 
The transition rates satisfy the normalization condition $\sum_{k,n} \gamma^\nu[k,n]=1$, reflecting the conservation 
of probability. 
Figure~\ref{fig:Opener}(b) illustrates quasiparticle transport processes.

Due to the AC Josephson effect, the oscillations in the Cooper pair current are shifted by the Josephson frequency 
$\omega_\text{J}=2eV/\hbar$, such that the current becomes a superposition of oscillatory components at frequencies 
$\omega_\text{J}+(E_n-E_m)/\hbar$. The time-averaged Cooper pair current is given by
\begin{subequations}\begin{align}
\overline{I}_{\text{cp}}=&\sum_{m,n} \delta_{eV, (E_m-E_n)/2} \, \overline{I}^{\text{cp}}_{mn},
\end{align}
where $\delta_{x,y}$ equals one for $x=y$, and zero otherwise. Therefore, the
transport of Cooper pairs results in spikes in the DC $I$-$V$
characteristics, occurring when the applied voltage satisfies the resonance
condition $eV=(E_m-E_n)/2$. For an overview, Fig.~\ref{fig:Opener}(c) shows the
spike structure due to Cooper pair transport, along with the step-like
quasiparticle current resulting from an $LC$-resonator used as the environment. 
Each pair of states $\ket{E_m}$, $\ket{E_n}$ contributes
\begin{align}
\overline{I}^{\text{cp}}_{mn}=&  \sum_k \Gamma_{\text{S}}(E_k-(E_m+E_n)/2) \Im \bigg ( e^{i \phi_0} C_{nkm} \bigg ),
\end{align}
with
\begin{align}
\Gamma_{\text{S}}(E)=& \, 2 \, \text{P.V.} \int \limits_{-\infty}^\infty d\epsilon \frac{\Gamma_{\text{cp}}(\epsilon)}{\pi(\epsilon-E)} 
\label{eq:SRate}\\
C_{nkm}=& \, (A^+_{kn})^*A^-_{km} \Tr \{\hat{L}_{km} \hat{\rho}\hat{L}_{kn}^\dagger \} \nonumber\\
=& \, (A^+_{kn})^*A^-_{km} \rho_{mn},
\label{eq:Ctensor}
\end{align}\end{subequations}
where $\text{P.V.}$ denotes principal value. The rate $\Gamma_{\text{S}}(eV)$ is related to the AC Josephson 
current $I_{c,1}(V)\sin(\phi_0+\omega_f t)$ by~\cite{Barone1982}
\begin{align}
    I_{c,1}(V)=&[\Gamma_{\text{S}}(eV) + \Gamma_{\text{S}}(-eV)] / 2.
    \label{eq:ACcurrRate}
\end{align}
The factor $C_{nkm}$ incorporates the effect of the environment, and contains the 
transition amplitudes $A^+_{kn}$ and $A^-_{km}$. We interpret this factor as an interference contribution between the 
transitions $E_m$ to $E_k$ and $E_n$ to $E_k$, with an initial coherence $\rho_{mn}$ being necessary to generate
interference between the transitions, see Fig.~\ref{fig:Opener}(b). Degenerate energy splittings among 
multiple levels lead to their collective contribution to a single spike in the $I$-$V$ characteristics. If all 
transition energies are distinct, the spike height is directly proportional to the coherences $\varrho_{mn}$ of 
the environmental density matrix, enabling their direct extraction. 

The physics at play here is closely related to the Shapiro effect that occurs under a harmonic current 
drive~\cite{Shapiro1963,Barone1982}. There, voltage plateaus emerge at multiples of half the driving frequency,
manifesting as current spikes in the voltage-source model~\cite{Barone1982}. Here, the spikes occur at the transition
energies of the environment and need not be regularly spaced.

Note that a system in equilibrium, or in an incoherent superposition of energy eigenstates, does not exhibit spikes 
at finite voltage. Interestingly, a modified supercurrent $\overline{I}^{\text{cp}}=\sum_{n,k} \Gamma_S(E_k-E_n) 
\Im (e^{i \phi_0} C_{nkn})$ arises even in equilibrium. This is in contrast to predictions from standard dynamical
Coulomb blockade theory~\cite{Ingold1992}, where the supercurrent vanishes and the first nonzero contribution to the
Cooper pair current is second order in the tunnel conductance. In Ref.~\cite{Ingold1992}, the Cooper pair condensate 
was modeled using the Josephson Hamiltonian. The microscopic analysis~\cite{Joyez2013} shows that the supercurrent
survives for environments in which the real part of the impedance vanishes as $\mathcal{O} (\omega^2)$ or faster as
$\omega \rightarrow 0$. A set of discrete modes satisfies this scaling behavior, and the supercurrent remains finite. 

\textit{Quantum harmonic oscillator.---} To illustrate the physics, we focus now on the example of a single 
electromagnetic mode modeled as a quantized $LC$ circuit. The flux $\hat{\Phi}=\hbar \hat{\varphi}/e$ through the
inductor is related to the phase operator $\hat{\varphi}=-i \sqrt{g}(\hat{a}-\hat{a}^\dagger)$,
with $g=\pi Z_{LC}/R_K$ the strength of the zero-point voltage fluctuations, determined by the impedance of the 
resonator $Z_{LC}=\sqrt{L/C}$ and the resistance quantum $R_{\text{K}}=2 \pi \hbar/e^2$, and 
$\hat{a}/\hat{a}^\dagger$ the annihilation/creation operator of the mode. The 
oscillator has eigenenergies $E_n=\hbar \omega_0 (n+1/2)$ and Fock states $\ket{n}$, with transition 
energies $E_{n+l}-E_n=l \hbar \omega_0, l \in \mathbb{Z}$ between the states,
where $\omega_0=1/\sqrt{LC}$ is the resonance frequency. To explore how different
quantum states influence the transport, we prepare the oscillator in either a
Fock superposition
\begin{subequations}\begin{align}
    \ket{\Psi_{01}}=\frac{\ket{0}+\exp(i\theta_1) \ket{1}}{\sqrt{2}},
\end{align}
a squeezed vacuum state
\begin{align}
\ket{\xi}= \sum^\infty_{n=0} (-1)^n \frac{\sqrt{(2n!)}}{2^n n!} \exp(in \theta_2) \frac{(\tanh(|\xi|))^{n}}{\sqrt{\cosh(|\xi|)}} \ket{2n},
\end{align}
or a coherent state
\begin{align}
    \ket{\alpha}=\exp \left(-\frac{|\alpha|^2}{2}\right) \sum_{n=0}^\infty \frac{|\alpha|^ne^{in\theta_3}}{\sqrt{n!}} \ket{n}.
\end{align}\label{eq:states}\end{subequations}
The squeezed state is controlled by the squeezing amplitude $|\xi|$, which characterizes the degree of squeezing, 
while the coherent state depends on $|\alpha|$, which determines the amplitude of the oscillations. All states can
carry an associated phase $\theta_s$ $(s=1,2,3)$. By simple inspection, we can infer from Eq.~\eqref{eq:Ctensor} 
and the properties of the states which Cooper pair current spikes may appear in the $I$-$V$ characteristics.
Since $\ket{\psi_{01}}$ only has nonzero coherences on the first off-diagonals of the density matrix, spikes 
only appear at $eV=-\hbar \omega_0/2,0,\hbar \omega_0/2$. A squeezed vacuum state possesses nonzero entries only 
for even occupation numbers, so no peaks appear at $eV=j\hbar \omega_0/2$, with $j$ odd. In contrast, a coherent 
state can have spikes at all $j\in \mathbb{Z}$ due to its nonzero coherences across all Fock states. The distinct
coherence structures of these states are directly reflected in the Cooper pair transport 
(see Fig.~\ref{fig:Opener}(c)).
\begin{figure}
    \centering
    \includegraphics[width=1\linewidth]{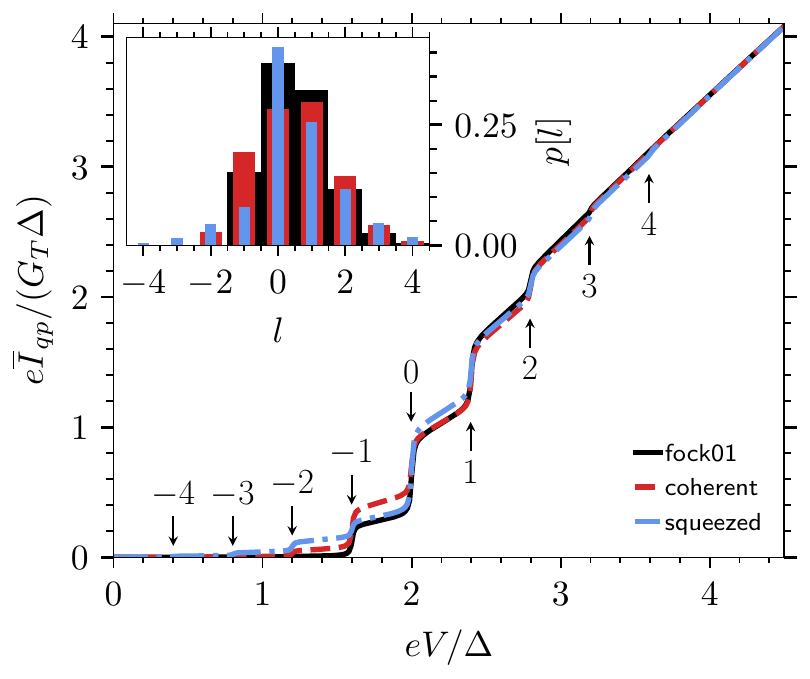}
    \caption{Quasiparticle current $\overline{I}_{\text{qp}}$ for the three different environmental states given in Eqs.~\eqref{eq:states}. The inset displays the probability $p[l]$ for exchanging $l$ energy quanta with the environment, while the arrows
    indicate the positions $2\Delta+l\hbar \omega_0$, where the $l$-th energy exchange 
    begins to contribute to the current. The parameters controlling the states are set to $|\xi|=|\alpha|=1$, $\theta_s \in \mathbb{R}$. The oscillator is characterized by $\hbar\omega_0=0.4\Delta$, $g=0.5$, and the tunnel rate \eqref{eq:QPTunnelrate} is evaluated at zero temperature.
    }
    \label{fig:QPcurrent}
\end{figure}

The background current in Fig.~\ref{fig:Opener}(c) arises from quasiparticle transport and is described by
\begin{subequations}\begin{align}
\overline{I}_{\text{qp}}=& \sum_{l=-\infty}^\infty   p[l] 
\sum_{\nu=\pm} \nu \Gamma_{\text{qp}}(l \hbar \omega_0 - \nu eV),
\label{eq:QPOsci}
\end{align}
where the quasiparticles exchange $l$ energy quanta with the environment, with probability
\begin{align}
    p[l]=\sum_{n=0}^\infty \Theta(n+l) |A^{-}_{n+ln}|^2 \rho_{nn},
\end{align}
determined by the transition amplitudes
\begin{align}
A^\nu_{n+ln}=&e^{-g/2} \begin{cases}
\sqrt{\frac{n!}{(n+l)!}}  (\nu g)^{l/2} L_n^{l}(g) & l \geq 0 \\
\sqrt{\frac{(n+l)!}{n!}} (-\nu g)^{-l/2} L_{n+l}^{-l}(g) & l<0
\end{cases} .
\end{align}\end{subequations}
Here, $L^l_n(x)$ are generalized Laguerre polynomials. 

Figure \ref{fig:QPcurrent} shows the quasiparticle current for the three environmental states in Eqs.~\eqref{eq:states}. 
Due to the superconducting energy gap, the quasiparticle rate $\Gamma_{\text{qp}}(E)$ becomes nonzero only for 
$E\leq -2\Delta$. Therefore, the processes involving the exchange of $l$ energy quanta with the resonator set in 
successively at $2\Delta+l \hbar \omega_0$, where the exchange probability $p[l]$ determines the height of the step. 
For example, the Fock superposition can emit at most one excitation, implying $p[l]=0$ for $l<-1$. As a result, the 
first step appears at $eV=2\Delta-\hbar \omega_0$, whereas for the squeezed state, the first step occurs at 
$2\Delta-4\hbar\omega_0$. The successive appearance of the steps enables direct extraction of the exchange 
probability $p[l]$ from the quasiparticle current. In contrast, for normal-conducting junctions, determining the 
exchange probability requires both the average current and current noise~\cite{Souquet2014}. As proposed by 
Souquet \emph{et al.}~\cite{Souquet2014}, the exchange probability can serve as a diagnostic tool to distinguish 
classical from nonclassical states.

Cooper pair transport occurs at $eV=j\hbar \omega_0/2$, $j \in \mathbb{Z}$, and the height of the resulting current 
spike is given by
\begin{subequations}\begin{align}
\overline{I}^{\text{cp}}_j=&\sum_{l=-\infty}^\infty  \Gamma_{\text{S}}([l-j/2]\hbar \omega_0) \Im \bigg ( e^{i \phi_0} 
\mathcal{C}[l,j] \bigg ),
\end{align}
where
\begin{align}
\mathcal{C}[l,j]=&\sum_{n=0}^\infty \Theta(n+l)\Theta(n+j) A^-_{nn+l}A^-_{nn+j} \rho_{n+jn} .
\label{eq:Inferfac}
\end{align}
This expression reduces to a current-phase relation
\begin{align}
    \overline{I}^{\text{cp}}_j = & \, \mathcal{I}_j \sin(\phi_0-\vartheta_j),
    \label{eq:CPosci}
\end{align}\end{subequations}
with the amplitude $\mathcal{I}_j$ and phase shift $\vartheta_j$ depending on the environmental state, the coupling 
strength $\sqrt{g}$, and the resonator frequency $\omega_0$. The $j$-th spike is sensitive to the $j$-th diagonal 
of the density matrix $\rho_{n+jn}$. A $\varphi_0$-junction cannot be realized in this case, as the modified supercurrent $\overline{I}^\text{cp}_0$ depends only on the diagonal elements of the density matrix, which are real. This leads to $\vartheta_0=0$, implying a purely sinusoidal current-phase relation.

\begin{figure}
    \centering
    \includegraphics[width=1\linewidth]{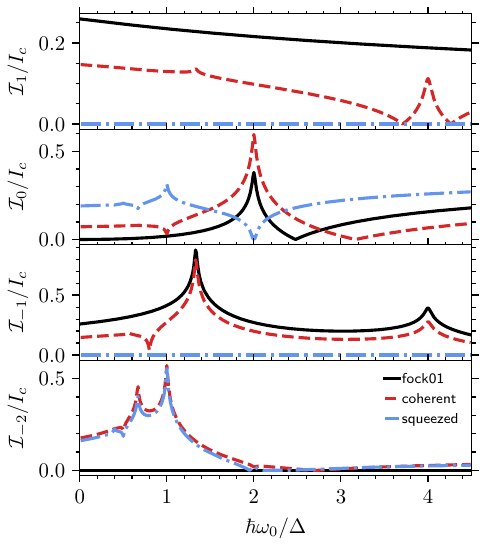}
    \caption{Cooper pair current amplitudes $\mathcal{I}_j$ (see Eq.~\eqref{eq:CPosci}) as a function of frequency 
    for the three different environmental states defined in Eqs.~\eqref{eq:states}. The amplitude 
    is normalized to the critical current $I_c$ of the junction in the absence of
    an environment, and the tunnel rate in Eq.~\eqref{eq:CPTunnelrate} is
    calculated at zero temperature. The state parameters are set to $|\xi|=|\alpha|=1$ with
    arbitrary phases $\theta_s \in \mathbb{R}$, while the coupling is $g=1/2$.}
    \label{fig:CPomega}
\end{figure}

Figure~\ref{fig:CPomega} shows the frequency 
dependence of the Cooper pair current amplitudes. The frequency dependence exhibits sharp features at 
specific frequencies, which are related to Riedel peaks, which are singularities at $|eV|=2\Delta$ in the AC 
Josephson current amplitude~\cite{Barone1982}. Here, the superconducting rate is 
related to the AC Josephson current amplitude through Eq.~\eqref{eq:ACcurrRate}. 
In contrast, the rate is asymmetric in 
energy and exhibits a Riedel peak only at $E=-2\Delta$~\cite{SupMat}. Hence, the frequency dependence of the 
amplitude $\mathcal{I}_j$ consists of replicas of the Riedel peak at $\hbar \omega_0=-4\Delta/(2l-j)$, weighted by 
the interference factors $\mathcal{C}[l,j]$. Kinks at $\mathcal{I}_j=0$ indicate a sign change of the spike and arise 
from taking the absolute value, $\mathcal{I}_j=|\overline{I}^{cp}_j|$. 

\begin{figure}
    \centering
    \includegraphics[width=1\linewidth]{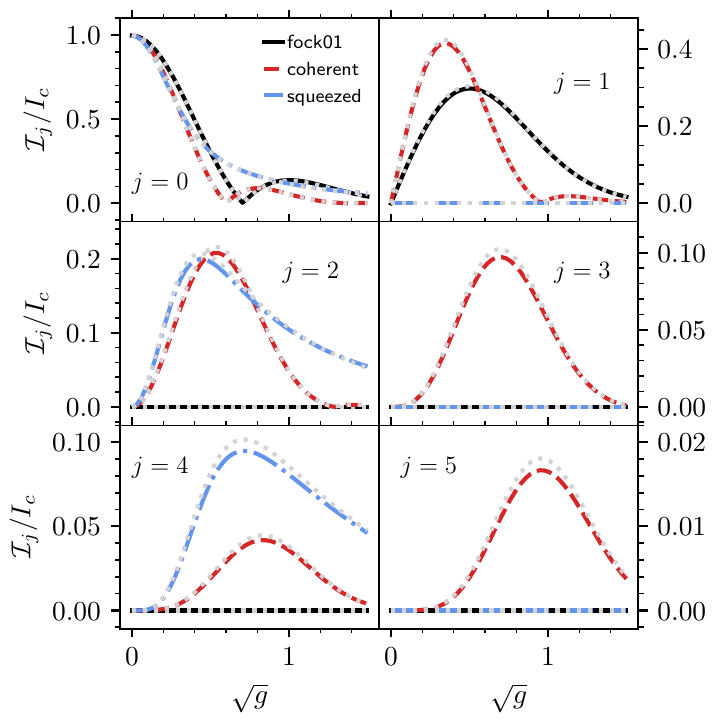}
    \caption{Cooper pair current amplitudes $\mathcal{I}_j$ (see Eq.~\eqref{eq:CPosci}) as a function of the coupling
    strength $\sqrt{g}$ for $\hbar \omega_0=0.1 \Delta$. The bare critical current $I_c$ is used to
    normalize the amplitudes, and the tunnel rate in Eq.~\eqref{eq:CPTunnelrate} is calculated at zero temperature.
    The amplitudes are evaluated for the three environmental states defined in Eqs.~\eqref{eq:states}, where each 
    $\mathcal{I}_j$ probes the $j$-th diagonal of the corresponding density matrix. The state parameters are given 
    by $|\xi|=|\alpha|=1$ with arbitrary phases $\theta_s \in \mathbb{R}$. The gray dashed lines indicate the 
    zero-frequency limit, which depends solely on the transition amplitudes $\tilde{A}^-_{nn+j}$ and the density 
    matrix elements $\rho_{n+jn}$.}
    \label{fig:Cpcoupling}
\end{figure}

Figure~\ref{fig:Cpcoupling} shows the coupling dependence of the amplitudes
$\mathcal{I}_j$ for $\hbar \omega_0=0.1\Delta$, and compares it to the zero-frequency limit (gray dashed lines). Varying the coupling strength modifies the transition amplitudes and, consequently, the interference factor~\eqref{eq:Inferfac}. The interference factor satisfies the identity $\sum_{l,j} \mathcal{C}[l,j]=\langle e^{2 \sqrt{g} 
(\hat{a}-\hat{a}^\dagger)} \rangle \equiv M[2 \sqrt{g}]$, which corresponds to the moment-generating function of 
the quadrature $\hat{a}-\hat{a}^\dagger$ evaluated at $2 \sqrt{g}$. As $\omega_0 \rightarrow 0$, the total Cooper 
pair current reduces to an expression involving the moment-generating function, $\sum_j \overline{I}^{\text{cp}}_j=I_c 
\Im(e^{i \phi_0} M[2 \sqrt{g}])$, where $I_c=\Gamma_{\text{S}}(0)$ is the critical current of the uncoupled junction. 
The individual amplitudes take the form $\lim_{\omega_0 \rightarrow 0}\mathcal{I}_j = I_c | \sum_n 
\Theta(n+j)\tilde{A}^-_{nn+j} \rho_{n+jn}|$, with a rescaled transition amplitude $\tilde{A}^-_{nn+j} = 
\bra{n}\exp(2i\hat{\varphi})\ket{n+j}$. 
Thus, the coupling dependence serves as a sensitive indicator of the environmental state.

\textit{Conclusions.---} We have shown that the coupling between a nonclassical electromagnetic environment and a 
Josephson tunnel junction gives rise to a very rich phenomenology in the current-voltage characteristics. In particular, 
we have illustrated how a superconducting tunnel junction can be used to completely characterize the nature of quantum 
states generated with circuit-QED setups, something that is not possible with normal tunnel junctions. Moreover, with 
the theoretical framework used here, the analysis of the transport characteristics can be straightforwardly extended 
to the study of higher-order cumulants like shot noise.

\begin{acknowledgments}
We sincerely thank Bertrand Reulet and David Christian Ohnmacht for helpful discussions.
This research was supported by the Deutsche Forschungsgemeinschaft (DFG; German Research 
Foundation) via SFB 1432 (Project \mbox{No.} 425217212), which in particular sponsored the 
stay of J.C.C.\ at the University of Konstanz as a Mercator Fellow.
\end{acknowledgments}

\section{Appendix: Hamiltonian}
Our objective is to analyze the impact of an arbitrary electromagnetic environment on the current through a superconducting tunnel junction. In the absence of an environment, quantum transport in the mesoscopic conductor is governed by the Hamiltonian $\hat{H}_{\text{cond}}$~\cite{Ingold1992,Levitov1996,Belzig2003,Snyman2008}. We model the coupling between the junction and the environment via the interaction Hamiltonian
\begin{subequations}\begin{align}
    \hat{H}_{\text{int}}=\frac{\hbar}{e}\hat{\phi}(t) \hat{I},
\end{align}
where $\hat{I}$ denotes the junction's current operator, $\hat{\phi}(t)$ represents the phase operator associated with the electromagnetic environment, and $e$ corresponds to the elementary charge. As a specific example, a constant phase operator corresponds to a static phase bias, which gives rise to the standard supercurrent. A classical, time-dependent voltage bias $V(t)$ induces a dynamic phase $\hat{\phi}(t)=\int^t_{t_{\text{s}}} dt' eV(t')/\hbar$, which describes the AC Josephson effect in the case of a constant voltage (see Eq.~\eqref{eq:ACJoseph}), and gives rise to Shapiro steps when the drive is harmonic. In general, the dynamics of the phase operator is captured by a Hamiltonian $\hat{H}_{\text{env}}$. Then, the total Hamiltonian is given by \cite{Nazarov2012}
\begin{align}
    \hat{H}=\hat{H}_{\text{cond}}+\hat{H}_{\text{env}}+\hat{H}_{\text{int}}.
    \label{eq:Hamil}
\end{align}\end{subequations}

\section{Appendix: Cumulant-generating functional}
The statistics of the current $\hat{I}_{\text{H}}(t)$ is fully encoded in the partition functional~\cite{Levitov1996,Belzig2003,Kindermann2004,Snyman2008,Nazarov2007} 
\begin{align}
\mathcal{Z}[\chi]= \bigg \langle \tilde{T}[e^{\frac{i}{e} \int_{t_{\text{s}}}^{\infty} dt \chi^-(t) \hat{I}_{\text{H}}(t)}] T[e^{-\frac{i}{e} \int_{t_{\text{s}}}^{\infty} dt \chi^+(t) \hat{I}_{\text{H}}(t)}] \bigg \rangle,
\label{eq:MomentF}
\end{align}
where $T$ and $\tilde{T}$ denote the time-ordering and anti-time-ordering operators, respectively. The functions $\chi^+(t)$ and $\chi^-(t)$ are the forward and backward components of the counting field defined on the Keldysh contour $C=C^++C^-$, which runs from $t_{\text{s}}$ to infinity and back to $t_{\text{s}}$. The subscript H indicates that the operator evolves in the Heisenberg picture, i.e., according to the Hamiltonian given in Eq. \eqref{eq:Hamil}. The average $\langle \cdot \rangle$ is taken with respect to the initial density matrix $\hat{\varrho}=\hat{\varrho}_{\text{cond}} \otimes \hat{\varrho}_{\text{env}}$, which is assumed to be separable. Taking the logarithm of the partition functional results in the cumulant-generating functional $\mathcal{S}[\chi]=\log(\mathcal{Z}[\chi])$. Time-ordered current correlations can be obtained by functionally differentiating the moment-generating functional with respect to the counting field:
\begin{widetext}\begin{subequations}\begin{align}
    \langle \tilde{T}[\hat{I}_{\text{H}}(t_1)\ldots\hat{I}_{\text{H}}(t_j)]T[\hat{I}_{\text{H}}(t_{j+1})\ldots \hat{I}_{\text{H}}(t_n)]\rangle=(-1)^j(ie)^n\frac{\delta^n \mathcal{Z}[\chi]}{\delta \chi^-(t_1)\ldots \delta \chi^-(t_j)\delta \chi^+(t_{j+1})\ldots \delta \chi^+(t_n)}\bigg |_{\chi=0} \\
    \langle \langle \tilde{T}[\hat{I}_{\text{H}}(t_1)\ldots\hat{I}_{\text{H}}(t_j)]T[\hat{I}_{\text{H}}(t_{j+1})\ldots \hat{I}_{\text{H}}(t_n)]\rangle \rangle=(-1)^j(ie)^n\frac{\delta^n \mathcal{S}[\chi]}{\delta \chi^-(t_1)\ldots \delta \chi^-(t_j)\delta \chi^+(t_{j+1})\ldots \delta \chi^+(t_n)} \bigg |_{\chi=0},
\end{align}\end{subequations}\end{widetext}
where $\langle \langle \cdot \rangle \rangle$ stands for the cumulant averaging. The average current and the noise are obtained by
\begin{subequations}\begin{align}
    \langle \hat{I}_{\text{H}}(t) \rangle =&ie \frac{\delta \mathcal{S}[\chi]}{\delta \chi^q(t)} \bigg |_{\chi=0} \label{eq:Curr}\\
    \langle \{\Delta \hat{I}_{\text{H}}(t),\Delta \hat{I}_{\text{H}}(t')\}\rangle=&-e^2\frac{\delta^2 \mathcal{S}[\chi]}{\delta \chi^q(t)\delta \chi^q(t')}\bigg |_{\chi=0},
\end{align}\end{subequations}
with the quantum field $\chi^q(t)=\chi^+(t)-\chi^-(t)$, $\{\cdot,\cdot\}$ the anti-commutator and $\Delta \hat{I}_{\text{H}}(t)=\hat{I}_{\text{H}}(t)-\langle \hat{I}_{\text{H}}(t)\rangle$ the difference to the mean~\cite{Kindermann2003,Blanter2000}. \\

The central idea is to express the partition functional \eqref{eq:MomentF} in terms of the known partition functional for an isolated tunnel junction~\cite{Nazarov2007,Snyman2008}. Accordingly, we transition to the interaction picture to decouple the dynamics of the junction and the environment. An operator $\hat{O}_{\text{H}}(t)$ in the Heisenberg picture is related to the operator in the interaction picture $\hat{O}_{\text{I}}(t)$ by 
\begin{subequations}
\begin{align}
\hat{O}_{\text{H}}(t)=\hat{U}_{\text{int}}^\dagger(t,t_{\text{s}})\hat{O}_{\text{I}}(t) \hat{U}_{\text{int}}(t,t_{\text{s}}),
\end{align}
where
\begin{align}
\hat{U}_{\text{int}}(t,t_{\text{s}})= T \exp \left(-\frac{i}{\hbar} \int_{t_{\text{s}}}^t \hat{H}^{\text{I}}_{\text{int}}(t') dt' \right),
\end{align}
is the interaction-picture evolution operator, and
\begin{align}
\hat{H}^{\text{I}}_{\text{int}}(t)=\frac{\hbar}{e}\hat{\phi}_{I}(t) \hat{I}_{I}(t),
\end{align}
\end{subequations}
with the subscript $I$ indicating operators in the interaction picture. The operator $\hat{\phi}_{\text{I}}(t)$ evolves under the Hamiltonian $\hat{H}_{\text{env}}$, while $\hat{I}_{\text{I}}(t)$ evolves under $\hat{H}_{\text{cond}}$. To simplify the notation, we promote the time variable to the full Keldysh contour $C$, and introduce the contour-ordering operator $T_{\text{C}}$, the contour evolution operator $\hat{U}_{\text{int}}^{\text{C}}=T_{\text{C}}\exp(-\frac{i}{\hbar}\int_{\text{C}} dt \hat{H}_{\text{int}}^{\text{I}} (t))$, and a unified counting field $\chi(t)$, which equals $\chi^+(t)$ on the forward branch and $\chi^-(t)$ on the backward branch of the contour. The partition functional transforms to
\begin{widetext}\begin{align}
\mathcal{Z}[\chi]=&\sum_{n=0}^\infty \frac{(-i)^n}{n!} \int_{\text{C}} dt_1 \ldots \int_{\text{C}} dt_n \langle T_{\text{C}}[ \chi(t_1) \hat{I}_{\text{H}}(t_1)\ldots \chi(t_n) \hat{I}_{\text{H}}(t_n)] \rangle \nonumber \\
=&1+\sum_{n=1}^\infty \frac{(-i)^n}{n!} \int_{\text{C}} dt_1 \ldots \int_{\text{C}} dt_n \langle T_{\text{C}}[ \hat{U}_{\text{int}}(t_{\text{s}},t_1)\chi(t_1) \hat{I}_{\text{I}}(t_1)\hat{U}_{\text{int}}(t_1,t_2)\ldots \hat{U}_{\text{int}}(t_{n-1},t_n)\chi(t_n) \hat{I}_{\text{I}}(t_n) \hat{U}_{\text{int}}(t_n,t_{\text{s}}) ]\rangle \nonumber\\
=&1+\sum_{n=1}^\infty \frac{(-i)^n}{n!} \int_{\text{C}} dt_1 \ldots \int_{\text{C}} dt_n \langle T_{\text{C}} [\hat{U}^{\text{C}}_{\text{int}}\chi(t_1) \hat{I}_{\text{I}}(t_1)\ldots \chi(t_n) \hat{I}_{\text{I}}(t_n)]\rangle+\langle T_{\text{C}} \hat{U}^{\text{C}}_{\text{int}}\rangle-\langle T_{\text{C}} \hat{U}^{\text{C}}_{\text{int}}\rangle \nonumber \\
=&1+\bigg \langle T_{\text{C}} \exp \left(-\frac{i}{e} \int_{\text{C}} [\chi(t)+\hat{\phi}_{\text{I}}(t)] \hat{I}_{\text{I}}(t)dt\right)  \bigg \rangle-\langle T_{\text{C}} \hat{U}^{\text{C}}_{\text{int}} \rangle \nonumber\\
=&1+\sum_{m=0}^\infty \frac{g^m}{m!} \int_{\text{C}} dt_1 \ldots \int_{\text{C}} dt_m \bigg [ \frac{\delta^m \mathcal{Z}_0[\chi]}{\delta  \chi(t_1) \ldots \delta \chi(t_m)}-\frac{\delta^m \mathcal{Z}_0[\chi]}{\delta \chi(t_1) \ldots \delta  \chi(t_m)} \bigg |_{\chi=0} \bigg ] \bigg \langle T_{\text{C}} [\hat{\phi}_{\text{I}}(t_1) \ldots \hat{\phi}_{\text{I}}(t_m)] \bigg  \rangle_{\text{env}},
\label{eq:MinM0}
\end{align}\end{widetext}
with the partition functional $\mathcal{Z}_0[\chi]$ of the isolated tunnel junction, and $\langle \cdot \rangle_{\text{env}}$ the average with respect to $\hat{\rho}_{\text{env}}$. The final line is obtained by expanding $\hat{U}_{\text{int}}^{\text{C}}$ and exploitation of the defining property of the partition functional. The term $\langle T_{\text{C}} \hat{U}^{\text{C}}_{\text{int}}\rangle$ ensures proper normalization of the partition functional, such that $\mathcal{Z}[0]=1$. The partition functional formally resembles a functional Taylor expansion of $1+\langle T_{\text{C}} [\mathcal{Z}_0[\chi+\delta \chi]-\mathcal{Z}_0[\delta \chi]]\rangle_{\text{env}}$, with a displacement $\delta \chi(t)=\hat{\phi}_{\text{I}}(t)$.\\

The cumulant-generating functional (or Keldysh action) for a general multiterminal and time-dependent scatterer was derived by Snyman and Nazarov~\cite{Snyman2008,Nazarov2015}. In a two-terminal device with an energy-independent scattering matrix, the cumulant-generating functional is given by~\cite{Belzig2001,Nazarov2007,Snyman2008,Nazarov2015}
\begin{align}
\mathcal{S}_0[\chi]=\frac{1}{4} \sum_n  \Tr \ln \bigg (1+\frac{T_n}{4} (\{\check{G}_1(\chi),\check{G}_2\}-2) \bigg ),
\end{align}
where $\check{G}_1(\chi)$ is the counting-field-dependent Green's function of terminal $1$, and $\check{G}_2$ the corresponding Green's function of terminal $2$, and $T_n$ are transmission eigenvalues. The Green's functions are treated as operators in time, Keldysh, Nambu, spin, and channel space, where $\log()$ denotes the operator logarithm, and $\Tr$ represents the trace over all these degrees of freedom. Operator multiplication involves integration over time variable, while the remaining spaces (Keldysh, Nambu, spin, and channel) entail summation over discrete indices. For example, $[\check{G}_1\check{G}_2](t,t'')\equiv\int_{\text{C}} dt' \check{G}_1(t,t') \check{G}_2(t',t'')$. Note that the the Green's functions $\check{G}_{1,2}$ are defined with a third Pauli matrix in Keldysh space to ensure normalization $\check{G}_{1,2}\check{G}_{1,2}=\check{1}$, but therefore the Keldysh $-+$/$--$ components correspond to the negative of the greater/anti-time ordered Green's function. The counting-field is incorporated in terminal $1$ by~\cite{Belzig2003,Kindermann2004,Nazarov2007}
\begin{align}
    \check{G}_1(\chi,t,t')=e^{-i \chi(t)\hat{\sigma}_3} \check{G}_1(t,t')e^{i \chi(t')\hat{\sigma}_3},
\end{align}
with $\hat{\sigma}_3$ the third Pauli-matrix in Nambu space.

We calculate the CGF $\mathcal{S}[\chi]$ in the tunnel limit and therefore retain only terms that are first-order in the tunnel conductance $G_T= e^2 \sum_n T_n/(\pi \hbar)$. In principle, higher-order terms can be obtained by expanding the CGF to the desired order in the tunnel conductance $G_T$, although the expressions become increasingly cumbersome. After tracing out the Keldysh and Nambu structure, we obtain
\begin{subequations}\begin{align}
\mathcal{S}[\chi]=&\frac{G_T\pi \hbar}{8e^2} \iint_{t_{\text{s}}}^{\infty} dt dt' \sum_{\substack{\nu,\nu'=\pm \\ \mu,\mu'=\pm}} (e^{-i \mu \chi^\nu(t)+i \mu' \chi^{\nu'}(t')}-1) \nonumber \\
&\times G_{\text{env}}^{\nu \nu' \mu \mu'}(t,t') \Tr \{ \tilde{G}_{1}^{\nu \nu' \mu \mu'}(t,t')  \tilde{G}_2^{\nu' \nu  \mu' \mu}(t',t)\},
\label{eq:CGF}
\end{align}
with $\tilde{G}_j^{\nu \nu' \mu \mu'},j=1,2$ a matrix in spin space, and $\Tr\{ \cdot \}$ the trace in spin space. The indices $\nu,\nu'$ refer to the Keldysh structure, with $\nu=+$ labeling the forward contour and $\nu=-$ the backward contour. The indices $\mu,\mu'$ correspond to the Nambu structure, where $\mu=+$ denotes the particle component and $\mu=-$ the hole component. The Green's functions $\check{G}_j$ are assumed to have a trivial spin structure, corresponding to the identity matrix. The environmental Green's functions are
\begin{align}
G^{++ \mu \mu'}_{\text{env}}(t,t')=&\langle T e^{-i \mu \hat{\phi}_{\text{I}}(t)} e^{i \mu'\hat{\phi}_{\text{I}}(t')} \rangle_{\text{env}}\\
G^{-- \mu \mu'}_{\text{env}}(t,t')=&\langle \tilde{T} e^{-i \mu \hat{\phi}_{\text{I}}(t)} e^{i \mu'\hat{\phi}_{\text{I}}(t')} \rangle_{\text{env}}\\
G^{+- \mu \mu'}_{\text{env}}(t,t')=&\langle  e^{i \mu'\hat{\phi}_{\text{I}}(t')} e^{-i \mu \hat{\phi}_{\text{I}}(t)} \rangle_{\text{env}}\\
G^{-+ \mu \mu'}_{\text{env}}(t,t')=&\langle  e^{-i \mu \hat{\phi}_{\text{I}}(t)} e^{i \mu' \hat{\phi}_{\text{I}}(t')} \rangle_{\text{env}}.
\end{align}\end{subequations}
\section{Appendix: Full counting statistics}
The full counting statistics (FCS) is obtained using the fields $\chi^{\pm}(t)=\pm (1-\Theta(t-t_{\text{e}}))\chi/2$, where $\chi$ is the counting variable and the measurement interval runs from $t_{\text{s}}$ to $t_{\text{e}}$. The cumulant generating function, derived from Eq.~\eqref{eq:CGF}, is given by
\begin{align}
    \mathcal{S}(\chi)=(e^{-i\chi}-1)N_{21}+(e^{i\chi}-1)N_{12}.
\end{align}
For a normal-metal junction, $N_{21}$ and $N_{12}$ represent the number of charges transferred from terminal $1$ to terminal $2$, and from terminal $2$ to terminal $1$, respectively. The charge transfer statistics follow a generalized Poisson distribution, as expected for transport through a tunnel junction \cite{Belzig2003}. Accounting for higher-order terms in the conductance $G_T$ causes deviations from Poissonian statistics, whereas in the absence of a quantum environment, the resulting full counting statistics is multinomial \cite{Belzig2001,Cuevas_Belzig_2003,Cuevas_Belzig_2004,Vanevic2008,Vanevic2007,Kindermann2003,Tobiska2005,Huebler2023}. A superconducting tunnel junction can exhibit negative values for $N_{21}$ and $N_{12}$, thereby undermining the interpretation of the FCS as a probability distribution. However, the FCS is observable when the dynamics of an idealized measurement device is taken into account~\cite{Belzig2001}. Quasiparticles and Cooper pairs contribute to charge transfer such that
\begin{subequations}\begin{align}
 N_{jk}=N_{jk}^{\text{qp}}+N_{jk}^{\text{cp}},   
\end{align}
with the individual contributions 
\begin{align}
N_{21}^{\text{qp}}=&\frac{1}{e}\int_{-\infty}^\infty dE   \sum_{\substack{\nu,\nu'=\pm\\ \nu \neq \nu'}} \overline{G}^{\nu \nu' \nu \nu}_{\text{env}}(E) \Gamma^{\nu \nu' \nu \nu}(E)\\
N_{21}^{\text{cp}}=&\frac{1}{e} \int_{-\infty}^\infty dE   \sum_{\substack{\nu,\nu'=\pm\\ \nu \neq \nu'}} \overline{G}^{\nu \nu \nu \nu'}_{\text{env}}(E) \Gamma^{\nu \nu \nu \nu'}(E)\\
N_{12}^{\text{qp}}=& \frac{1}{e}\int_{-\infty}^\infty dE   \sum_{\substack{\nu,\nu'=\pm\\ \nu \neq \nu'}} \overline{G}^{\nu' \nu \nu \nu}_{\text{env}}(E) \Gamma^{\nu' \nu \nu \nu}(E)\\
N_{12}^{\text{cp}}=& \frac{1}{e} \int_{-\infty}^\infty dE   \sum_{\substack{\nu,\nu'=\pm\\ \nu \neq \nu'}} \overline{G}^{\nu \nu \nu' \nu}_{\text{env}}(E) \Gamma^{\nu \nu \nu' \nu}(E),
\end{align}
where 
\begin{align}
    \overline{G}_{\text{env}}^{\nu \nu' \mu \mu'}(E)=&\iint_{t_{\text{s}}}^{t_{\text{e}}} \frac{dt dt'}{4 \pi \hbar} e^{iE(t-t')/\hbar} G^{\nu \nu' \mu \mu'}_{\text{env}}(t,t')
\end{align}
is the time-averaged environmental Green's function. In the absence of an electromagnetic environment, the tunneling rates are
\begin{align}
\Gamma^{\nu \nu'\mu \mu'}(E)=\frac{G_T}{8 e} \int_{-\infty}^\infty d \epsilon \Tr \{ \tilde{G}_1^{\nu \nu'\mu \mu'}(\epsilon-E) \tilde{G}_2^{\nu' \nu \mu' \mu}(\epsilon) \},
\label{eq:Rates}
\end{align}\end{subequations}
with the energy domain representation $G^{\nu \nu' \mu \mu'}_j(E)=\int_{-\infty}^\infty e^{iE\tau/\hbar} d\tau G^{\nu\nu'\mu\mu'}_j(\tau)$, where $\tau=t-t'$. Here, the terminals are assumed to be in thermal equilibrium, which implies time-translation invariance of the Green's functions, i.e., $G^{\nu \nu' \mu \mu'}_j(t,t')=G^{\nu \nu' \mu \mu'}_j(t-t')$. The Green’s function in Keldysh space takes the form:
\begin{align}
    \check{G}_j=& \begin{pmatrix}
\hat{G}_j^{++} & \hat{G}_j^{+-} \\
\hat{G}_j^{-+} & \hat{G}_j^{--}
\end{pmatrix}\nonumber\\
=& \frac{1}{2} \begin{pmatrix}
\hat{G}_j^a+\hat{G}_j^r+\hat{G}_j^k & \hat{G}_j^a-\hat{G}_j^r+\hat{G}_j^k \\
\hat{G}_j^a-\hat{G}_j^r-\hat{G}_j^k & \hat{G}_j^a+\hat{G}_j^r-\hat{G}_j^k
\end{pmatrix},
\end{align}
with the retarded/advanced $\hat{G}_j^{r/a}(E)$ component, and the Keldysh component $\hat{G}_j^k=(\hat{G}_j^r(E)-\hat{G}_j^a(E))(1-2f_j(E))$, where $f_j(E)=(1+e^{(E-\tilde{\mu}_j)/(k_\text{B}T_j)})^{-1}$ is the Fermi distribution. The Fermi distribution is determined by the temperature $T_j$, and the chemical potential $\tilde{\mu}_j$ of the corresponding terminal.\\
A pure voltage bias $V$ results in the environmental Green's functions 
\begin{align}
   \overline{G}_{\text{env}}^{\nu \nu' \mu \mu'}(E)=&\frac{t_0\delta(E-\mu' eV)e^{i\phi_0 (\mu'-\mu)/2}}{2} \nonumber\\
   &\cdot \begin{cases}
    1  & \mu =\mu' \\
    \frac{ \exp \left(i\mu'\omega_J t_0\right)-1}{i\mu'\omega_Jt_0} & \mu \neq \mu'
   \end{cases},
\end{align}
with Josephson frequency $\omega_J=2eV/\hbar$, the superconducting phase difference $\phi_0$, and the length of the measurement interval $t_0=t_{\text{e}}-t_{\text{s}}$. Here, we assume that the measurement interval is large, such that $\omega_J t_0 \gg 1$. The components $N_{jk}^{\text{cp}}$ oscillate at $\omega_J$, reflecting the AC Josephson effect. As expected for transport under voltage bias \cite{Belzig2003}, the number of transferred quasiparticles is given by the tunneling rates times the measurement interval.

\section{Appendix: Tunnel current}
The tunnel current $I(t)=\langle \hat{I}_{\text{H}}(t)\rangle$ is obtained by functionally differentiating the CGF with respect to the counting field (see Eq.~\eqref{eq:Curr}). This yields the expression 
\begin{align}
I(t)=\sum_{\substack{\nu \neq \nu' =\pm \\ \mu \mu'=\pm}} \int_{-\infty}^\infty  dE  P^{\nu \nu'\mu \mu'}(t,E) \Gamma^{\nu \nu'\mu\mu'}(E),
\end{align}
with the generalized $P$-function
\begin{align}
P^{\nu \nu' \mu \mu'}(t,E)=& \int_0^{t-t_{\text{s}}} \frac{d\tau}{2 \pi \hbar} \bigg (\mu \nu e^{iE \tau/\hbar} G_{\text{env}}^{\nu \nu' \mu \mu'}(t,t-\tau) \nonumber \\
&- \mu' \nu' e^{-iE \tau/\hbar} G^{\nu \nu'\mu \mu'}_{\text{env}}(t-\tau,t) \bigg ).
\end{align}
We employed the identities relating time-ordered and anti-time-ordered Green's functions to the lesser and greater components in order to eliminate the $++$/$--$ contributions. 

\section{Appendix: Normal conducting tunnel junction}
In this section, we examine a tunnel junction between two normal conductors interacting with an environment. The retarded/advanced Green's function of a normal lead is
\begin{align}
\hat{G}^{r/a}(E)=\pm \hat{\tau}_0 \otimes \hat{\sigma}_3,
\end{align}
with
with $\hat{\tau}_0$ the zeroth Pauli matrix in spin space, and $\hat{\sigma}_3$ the third Pauli matrix in Nambu space. The rates $\Gamma^{\nu \nu' \mu \mu'}$ with $\mu \neq \mu'$ vanish. The remaining nonzero rates simplify to $\Gamma^{\nu \nu' \mu \mu}(E)=\Gamma(\nu' E)$, for $\nu \neq \nu'$, and $\Gamma(E)$ represents the tunneling rate of the uncoupled junction:
\begin{align}
\Gamma(E)=&\frac{G_T}{e} \int_{-\infty}^\infty d \epsilon [1- f(\epsilon-E)] f(\epsilon)\nonumber\\
=&\frac{G_T}{e} \frac{E}{1-e^{-E/(k_\text{B}T)}},
\end{align}
assuming equal temperatures $T_1=T_2=T$. $f(E)$ denotes the Fermi function evaluated at zero chemical potential. The average current reduces to
\begin{align}
I(t)=\sum_{\nu=\pm} \nu \int_{-\infty}^\infty dE P^\nu(t,E) \Gamma(E),
\end{align}
with the $P$-functions
\begin{align}
P^\nu(t,E)=&\int_{0}^{\infty}\hspace{-1pt} \frac{d \tau}{\pi \hbar} \Re \bigg (e^{iE \tau/\hbar}  \langle e^{i \nu \hat{\phi}_{\text{I}}(t)}e^{-i \nu \hat{\phi}_{\text{I}}(t-\tau)} \rangle  \bigg ),
\label{eq:Pfunc}
\end{align}
where, for brevity, $\langle \cdot \rangle \equiv \langle \cdot \rangle_{\text{env}}$. We consider the perturbation to have been switched on in the distant past, and took the limit $t_{\text{s}} \rightarrow - \infty$. Therefore, we recover the standard expressions of dynamical Coulomb blockade physics (DCB)~\cite{Ingold1992,Joyez2013}, and the formula derived in~\cite{Souquet2014} for a non-equilibrium environment.
\section{Appendix: Superconducting tunnel junction}
We turn to the case of a superconducting tunnel junction coupled to an environment, examining how the environment impacts both the quasi-particle and Cooper pair tunneling currents. The superconducting Green’s function acquires a nontrivial matrix structure in Nambu space, reflecting the intrinsic particle–hole coherence characteristic of the superconducting state. The retarded and advanced Green's functions are given by
\begin{align}
\hat{G}_{j}^{r/a}=\hat{\tau}_0 \hspace{-1pt}\otimes \hspace{-1pt} \frac{-i}{\sqrt{\Delta_{j}^2-(E\pm i \eta)^2}} \begin{pmatrix}
(E\pm i \eta) & \Delta_{j} e^{i \phi_{j}} \\
 -\Delta_{j} e^{-i\phi_{j}}  &  -(E \pm i \eta)
\end{pmatrix},
\end{align}
with $\hat{\tau}_0$ the identity in spin space, $\Delta_j$ the superconducting gap, $\phi_j$ the superconducting phase, and the regularisation parameter $\eta$, which tends to zero, but can be kept finite to describe inelastic processes, which broaden the electronic states. The superconducting phases $\phi_j$ are set to zero in the following, as they are already accounted for in $\hat{\phi}_{\text{I}}(t)$. The normalization condition requires $\hat{G}^{r/a}\hat{G}^{r/a}=\hat{1}$, where $\hat{1}$ is the identity in spin-Nambu space.\\
The quasi-particle current is associated with the diagonal components in Nambu space, corresponding to terms with $\mu=\mu'$. The corresponding rates are $\Gamma^{\nu \nu' \mu \mu}(E)=\Gamma_{\text{qp}}(\nu'E)$, $ \nu \neq \nu'$, with 
\begin{subequations}\begin{align}
\Gamma_{\text{qp}}(E)=&\frac{G_T}{e} \int_{-\infty}^\infty d \epsilon \varrho_1(\epsilon-E) \varrho_2(\epsilon)[1-f(\epsilon-E)]  f(\epsilon),
\end{align}
and the superconducting density of states
\begin{align}
\varrho_j(E)=\Re \left(\frac{-i(E+i\eta)}{\sqrt{\Delta^2_j-(E+i\eta)^2}} \right).
\end{align}\end{subequations}
The resulting quasi-particle current,
\begin{align}
I(t)=\sum_{\nu=\pm} \nu \int_{-\infty}^\infty dE P^\nu(t,E) \Gamma_{\text{qp}}(E),
\end{align}
exhibits the same structure as the electron current discussed in the previous section and involves the same $P$-function defined in Eq.~\eqref{eq:Pfunc}. The only difference lies in the bare quasi-particle tunneling rate, which additionally incorporates the superconducting densities of states.\\
The Cooper pair current originates from the anomalous (off-diagonal) components of the Green’s functions, corresponding to $\mu\neq \mu'$, and gives rise to transition rates of the form $\Gamma^{\nu \nu' \mu \mu'}(E)=- \Gamma_{\text{cp}}(\nu' E)$, for $\nu \neq \nu'$ and $\mu \neq \mu'$. The Cooper pair tunneling rate $\Gamma_{\text{cp}}(E)$ is given by
\begin{subequations}\begin{align}
\Gamma_{\text{cp}}(E)=&\frac{G_T}{e} \int_{-\infty}^\infty d \epsilon \varsigma_1(\epsilon-E)\varsigma_2(\epsilon)(1-f(\epsilon-E))  f(\epsilon),
\end{align}
with the pair densities
\begin{align}
    \varsigma_j(E)=\Re \left( \frac{i \Delta_j}{\sqrt{\Delta_j^2-(E+i \eta)^2}}\right).
\end{align}\end{subequations}
Cooper pair tunneling leads to the contribution
\begin{align}
I_{\text{cp}}(t)=&\sum_{\nu=\pm} \nu \int_{-\infty}^\infty dE C^\nu(t,E) \Gamma_{\text{cp}}(E), 
\end{align}
with a correlation function
\begin{align}
C^\nu(t,E)=& \int_{0}^{\infty} \frac{d \tau}{\pi \hbar} \Re \bigg (e^{iE\tau/\hbar}   \langle e^{-i \nu \hat{\phi}_{\text{I}}(t)}e^{-i \nu \hat{\phi}_{\text{I}}(t-\tau)}  \rangle \bigg ). 
\end{align} 
A dc voltage $V$ and a superconducting phase shift applied additionally to the tunnel junction leads to a dynamical phase $\varphi_{dc}(t)=eVt/\hbar+\phi_0/2$, and can be incorporated by $\hat{\phi}_{\text{I}}(t) =\hat{\varphi}_{\text{I}}(t)+\varphi_{dc}(t)$, where $\hat{\varphi}$ is the phase operator of the environment. Thus, the $P$- and $C$-function are
\begin{widetext}\begin{subequations}\begin{align}
P^\nu(t,E)=&\int_{0}^{\infty}\hspace{-1pt} \frac{d \tau}{\pi \hbar} \Re \bigg (e^{i(E+\nu eV) \tau/\hbar}  \langle e^{i \nu \hat{\varphi}_{\text{I}}(t)}e^{-i \nu \hat{\varphi}_{\text{I}}(t-\tau)} \rangle \bigg )\\
C^\nu(t,E)=& \int_{0}^{\infty} \frac{d \tau}{\pi \hbar} \Re \bigg (e^{i(E+ \nu eV)\tau/\hbar} e^{-i \nu \phi_0-2i\nu eV t/\hbar} \langle e^{-i \nu \hat{\varphi}_{\text{I}}(t)}e^{-i \nu \hat{\varphi}_{\text{I}}(t-\tau)}  \rangle \bigg ). 
\end{align}\end{subequations}\end{widetext}
To the best of our knowledge, the $C$-function has not been discussed in the literature in this level of generality and constitutes the novel part of our work.
\section{Appendix: AC-Josephson effect and Shapiro steps}
\begin{figure}
    \centering
    \includegraphics[width=1\linewidth]{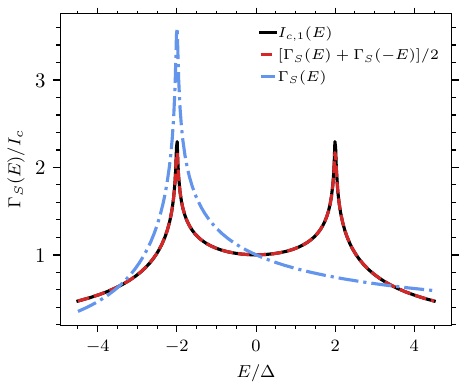}
    \caption{Comparison between the rate $\Gamma_{\text{S}}$ from Eq.~(6c) of the main text and the critical current $I_{\text{c},1}$. The plot illustrates that the critical current is given by the symmetrized rate evaluated at $\pm eV$.}
    \label{fig:Rate}
\end{figure}

A constant voltage bias represents the most basic example of an electromagnetic environment. The $C$-function reduces to
\begin{align}
    C^\nu(t,E)=&\cos(\phi_0+2eVt/\hbar)\delta(E+\nu eV) \nonumber\\
    &+\nu \sin(\phi_0+2eVt/\hbar) \frac{1}{\pi} P.V.\left[ \frac{1}{E+\nu eV}\right],
\end{align}
with $P.V.$ the principal value. The Cooper pair current is given by
\begin{subequations}\begin{align}
    I_{\text{cp}}(t)=&I_{\text{c},1}(V) \sin(\phi_0+2eVt/\hbar) \nonumber\\
    &+ I_{\text{c},2}(V) \cos(\phi_0+2eVt/\hbar),
    \label{eq:ACJoseph}
\end{align}
with the critical currents
\begin{align}
I_{\text{c},1}(V)=&\frac{G_T}{\pi e} P.V.\iint_{-\infty}^\infty dE d \epsilon \frac{\varsigma_1(E) \varsigma_2(\epsilon)}{E-\epsilon-eV}[f(E)-f(\epsilon)] \\
I_{\text{c},2}(V)=&\frac{G_T}{e}\int_{-\infty}^\infty d \epsilon \varsigma_1(\epsilon-eV) \varsigma_2(\epsilon)[f(\epsilon-eV)-f(\epsilon)]. 
\end{align}\end{subequations}
We recover the well-known AC Josephson effect~\cite{Josephson1962,Barone1982}. The rate $\Gamma_{\text{S}}(E)$, as defined in the main text, is related to the critical current amplitude $I_{\text{c},1}(V)$ via the relation $I_{\text{c},1}(V)=[\Gamma_{\text{S}}(eV)+\Gamma_{\text{S}}(-eV)]/2$, as shown in Fig.~\ref{fig:Rate}.
\\
An external microwave field incident at the tunnel junction is modeled by a voltage $V(t)=V_0 \cos(\omega_0 t)$, where $V_0$ and $\omega_0$ are the amplitude and angular frequency of the microwave drive, respectively. The time-dependent voltage induces a dynamical phase
\begin{align}
    \phi(t)=\frac{eV}{\hbar}t+\alpha\sin(\omega_0 t),
\end{align}
where $V$ corresponds to the applied dc voltage, and $\alpha=eV_0/(\hbar \omega_0)$ to the strength of the drive. Therefore, the environmental phase $\hat{\phi}(t)=\phi(t)$ results in
\begin{align}
C^\nu(t,E)=& \sum_{n,l=-\infty}^\infty J_n\left( \alpha \right)J_l\left( \alpha \right) \bigg (\cos(\phi_0+ \Omega_{nl}t) \nonumber\\
&\times \delta(E+\nu eV+\nu l \hbar \omega_0)+\nu \sin(\phi_0+ \Omega_{nl}t)\nonumber\\
&\times \frac{1}{\pi} P.V. \left[\frac{1}{E+\nu eV+\nu l \hbar \omega_0} \right] \bigg )
\end{align}
with $\Omega_{nl}=2eV/\hbar+(n+l) \omega_0$. We utilized the identity $\exp(iz\sin(\Theta))=\sum_{l=-\infty}^\infty J_l(z)\exp(il \Theta)$, where $J_l(z)$ denotes the Bessel function of the first kind of order $l$. The resulting Cooper pair current
\begin{align}
I_{\text{cp}}(t)=& \sum_{n,l=-\infty}^\infty J_n\left( \alpha \right)J_l\left( \alpha \right) [I_{\text{c},2}(eV+l \hbar \omega_0) \nonumber\\
\times& \cos(\phi_0+\Omega_{nl}t)+I_{\text{c},1}(eV+l \hbar \omega_0) \sin(\phi_0+\Omega_{nl}t) ]
\end{align}
oscillates at harmonics of the driving frequency and is consistent with the results reported in~\cite{Barone1982}. The time averaged Cooper pair current
\begin{align}
    \overline{I}_{\text{cp}}=&\lim_{T \rightarrow \infty}\frac{1}{T} \int_{-T/2}^{T/2} I_{\text{cp}}(t) dt \nonumber\\
    =& \sum_{m=-\infty}^\infty \delta_{eV,m\hbar \omega_0/2} \sum_{l=-\infty}^\infty J_{m-l}\left( \alpha \right)J_l\left( \alpha \right) \nonumber   \\
    \times& [I_{\text{c},1}(eV+l \hbar \omega_0) \sin(\phi_0)-I_{\text{c},2}(eV+l \hbar \omega_0) \cos(\phi_0)]
\end{align}
exhibits spikes at $eV=m\hbar \omega_0/2$ in the $I-V$ curve, with $\delta_{eV,m \hbar \omega_0/2}=1$ for $eV=m \hbar \omega_0/2$, and $\delta_{eV,m \hbar \omega_0/2}=0$ for $eV \neq m \hbar \omega_0/2$. These spikes correspond to Shapiro steps in a current-biased tunnel junction~\cite{Shapiro1963,Barone1982}.

\section{Appendix: State preparation}
We envisage a finite-time state preparation protocol that initializes the environment in a desired state~\cite{Hofheinz2009}, after which the average tunnel current is measured over a substantially longer time interval. The time-averaged current of interest is
\begin{align}
\overline{I}\equiv&\frac{1}{t_{\text{e}}-t_{\text{m}}} \int_{t_{\text{m}}}^{t_{\text{e}}} dt I(t)=\overline{I}_{in}+\overline{I}_{prep},
\end{align}
where the preparation takes place from time $t_{\text{s}}$ up to $t_{\text{m}}$, at which point the measurement begins and continues until $t_{\text{e}}$. The current $\overline{I}_{in}$ depends solely on the prepared state, while $\overline{I}_{prep}$ depends on the preparation protocol. The integrals in $\overline{I}_{in}$ run over times later than $t_{\text{m}}$ such that the environmental correlator can be reformulated as
\begin{align}
\langle e^{i \nu \hat{\varphi}_{\text{I}}(t)} e^{-i \nu' \hat{\varphi}_{\text{I}}(t')} \rangle=&\Tr \{U^\dagger(t,t_{\text{s}})e^{i \nu \hat{\varphi}}  U(t,t_{\text{s}})\nonumber\\
&\times U^\dagger(t',t_{\text{s}}) e^{-i \nu' \hat{\varphi}} U(t',t_{\text{s}})\nonumber\\
&\times \hat{\rho}_{\text{env}}\} \nonumber\\
=&\Tr \{ U^\dagger(t,t_{\text{m}})e^{i \nu \hat{\varphi}} U(t,t_{\text{m}}) \nonumber \\
&\times U^\dagger(t',t_{\text{m}}) e^{-i \nu' \hat{\varphi}} U(t',t_{\text{m}}) \nonumber\\
&\times U(t_{\text{m}},t_{\text{s}}) \hat{\rho}_{\text{env}}U^\dagger(t_{\text{m}},t_{\text{s}})\} \nonumber\\
=&\Tr \{ e^{i \nu \hat{\varphi}_{\text{m}}(t)}  e^{-i \nu' \hat{\varphi}_{\text{m}}(t')} \hat{\rho}_{in}\},
\end{align}
with $\hat{\rho}_{\text{env}}$ is the initial density matrix at $t_{\text{s}}$, $\hat{\rho}_{in}$ the initialized density matrix at $t_{\text{m}}$, and $U(t,t_{\text{s}})$ the time-evolution operator of the environment. The average current $\overline{I}_{prep}$ involves correlators of the form
\begin{align}
&\frac{1}{t_{\text{e}}-t_{\text{m}}} \int_0^{t_{\text{e}}-t_{\text{m}}} dt \int_{0}^{t_{\text{m}}-t_{\text{s}}}  \frac{d \tau}{\pi \hbar}\Re \bigg (e^{iE (t+\tau)/\hbar} \nonumber \\
&\times\langle e^{i \nu \hat{\varphi}_{\text{I}}(t+t_{\text{m}})}e^{-i \nu' \hat{\varphi}_{\text{I}}(t_{\text{m}}-\tau)}\rangle \bigg ),
\end{align}
and is sensitive to the time evolution of the environmental operator during the state preparation period. If, $\hat{H} \ket{E_n}=E_n \ket{E_n}$, with a discrete set of eigenvalues $E_n$ and orthonormal eigenstates $\ket{E_n}$, then
\begin{align}
    &\frac{1}{t_{\text{e}}-t_{\text{m}}} \int_0^{t_{\text{e}}-t_{\text{m}}} dt e^{iEt/\hbar}e^{i \nu \hat{\varphi}_{\text{m}}(t) }=\sum_{n,m}\bra{n}e^{i \nu \hat{\varphi}}\ket{m} \nonumber\\
    &\times \ket{n}\bra{m} \frac{1}{t_{\text{e}}-t_{\text{m}}} \int_0^{t_{\text{e}}-t_{\text{m}}}  dt e^{i(E+E_n-E_m)t/\hbar}. 
\end{align}
The time average leads to a factor
\begin{align}
    &\frac{1}{t_{\text{e}}-t_{\text{m}}}\int_0^{t_{\text{e}}-t_{\text{m}}}  dt e^{i(E+E_n-E_m)t/\hbar}= \frac{\cos(\Omega(t_{\text{e}}-t_{\text{m}}))}{i\Omega(t_{\text{e}}-t_{\text{m}})}\nonumber\\
    &+\frac{\sin(\Omega(t_{\text{e}}-t_{\text{m}}))}{\Omega(t_{\text{e}}-t_{\text{m}})}-\frac{1}{i\Omega(t_{\text{e}}-t_{\text{m}})}\xrightarrow[t_{\text{e}}-t_{\text{m}} \rightarrow \infty]{} \delta_{\Omega,0},
\end{align}
with $\Omega=(E+E_n-E_m)/\hbar$. In the limit of a large measurement window, this factor becomes non-zero only when $E+E_n-E_m=0$. Therefore, the integrand in the energy integral over $E$ is non-zero only on a set of measure zero, implying that $\overline{I}_{prep} \rightarrow 0$ in the limit $t_{\text{e}}-t_{\text{m}} \rightarrow \infty$. Including a finite voltage bias merely shifts $\Omega$, while the underlying reasoning remains unchanged. This demonstrates that the average current $\overline{I}=\overline{I}_{in}$ depends on the initialized state as well as on the system's evolution during the measurement interval, and not on the preparation protocol.

\section{Appendix: Example states}
The states examined in the main text are a Fock state superposition, a coherent state, and a squeezed vacuum state. Charge transport depends on these states through the density matrix $\rho_{n+jn}=\bra{n+j}\hat{\rho}\ket{n}$, $n+j, n \in \mathbb{N}$; see Eqs.~(8a),(9b) of the main text. The Fock state superposition $\ket{\psi_{01}}=[\ket{0}+e^{i\theta_1} \ket{1}]/2$ possesses the density matrix element
\begin{align}
    \rho_{n+jn}=\frac{1}{2}\begin{cases} 
    1 & n\leq 1 \, \text{and} \, j=0\\
    e^{i\theta_1} & n=0 \, \text{and} \, j= 1\\
    e^{-i\theta_1} & n=1 \, \text{and} \, j= -1\\
    \end{cases}.
\end{align}
The Coherent state $\ket{\alpha}$ yields a density matrix element
\begin{align}
    \rho_{n+jn}=\frac{|\alpha|^{2n} \alpha^{*j}}{\sqrt{n! (n+j)!}}e^{-|\alpha|^2},
\end{align}
with $\alpha=|\alpha|e^{i\theta_2} \in \mathbb{C}$ the coherent parameter. A squeezed vacuum state $\ket{\xi}$, characterized by the squeezing parameter $\xi=r e^{i \theta_3}$, with $0 \leq r < \infty, 0 \leq \theta <2 \pi$, is given by \cite{Gerry2023}
\begin{align}
\ket{\xi}=\frac{1}{\sqrt{\cosh(r)}} \sum^\infty_{m=0} (-1)^m \frac{\sqrt{(2m!)}}{2^m m!} e^{im \theta_3} (\tanh(r))^{m} \ket{2m}.
\end{align}
The density matrix elements reduce to
\begin{align}
\varrho_{n+jn}=&\begin{cases}
B_{nj}e^{i j \theta_3/2} \frac{[\tanh(r)]^{n+j/2}}{\cosh(r)}     & n,j \, \text{even} \\
0 & \text{else}
\end{cases},
\end{align}
with $B_{nj}=(-1)^{j/2}\sqrt{(n+j)!n!}/(2^{n+j/2} ((n+j)/2)!(n/2)!)$. The structure of the density matrix is directly reflected in the Cooper pair transport: for the Fock state, only the peaks at $j=-1,0,1$ appear; for the squeezed vacuum state, only even-$j$ peaks are present; while the coherent state exhibits peaks at all $j$.

\bibliography{Bib}

\end{document}